\documentclass[12pt,article]{aastex}

\usepackage{graphicx,natbib,textcomp,caption,subcaption}         
\usepackage[utf8]{inputenc}
\usepackage{xcolor,amsmath}
\usepackage{multirow}
\usepackage{float}

\slugcomment{ApJ manuscript, \today}

\shorttitle{Nature of Els\"{a}sser variables}
\shortauthors{Magyar, Van Doorsselaere, \& Goossens}

\begin{document}

\title{Understanding uniturbulence: self-cascade of MHD waves in the presence of inhomogeneities}

\author{N. Magyar}

\affil{Centre for Fusion, Space and Astrophysics, Physics Department, University of Warwick, Coventry CV4 7AL, UK; norbert.magyar@warwick.ac.uk;}
\affil{Centre for mathematical Plasma Astrophysics (CmPA), KU 
Leuven, Celestijnenlaan 
200B bus 2400, 3001 Leuven, Belgium;}

\author{T. Van Doorsselaere}
\affil{Centre for mathematical Plasma Astrophysics (CmPA), KU 
Leuven, Celestijnenlaan 
200B bus 2400, 3001 Leuven, Belgium;}
\author{M. Goossens}
\affil{Centre for mathematical Plasma Astrophysics (CmPA), KU 
Leuven, Celestijnenlaan 
200B bus 2400, 3001 Leuven, Belgium;}
\begin{abstract}

It is widely accepted in the MHD turbulence community that the nonlinear cascade of wave energy requires counter-propagating Alfv\'enic wave-packets, along some mean magnetic field. This fact is an obvious outcome of the MHD equations under the assumptions of incompressibility and homogeneity. Despite attempts to relax these assumptions in the context of MHD turbulence, the central idea of turbulence generation persists. However, once the assumptions of incompressiblity and homogeneity break down, the generally accepted picture of turbulent cascade generation is not universal. In this paper, we show that perpendicular inhomogeneities (across the mean magnetic field) lead to propagating wave solutions which are necessarily described by co-propagating Els\"{a}sser fields, already in the incompressible case. One simple example of these wave solutions is the surface Alfv\'en wave on a planar discontinuity across the magnetic field. We show through numerical simulations how the nonlinear self-deformation of these unidirectionally propagating waves leads to a cascade of wave energy across the magnetic field. The existence of this type of unidirectional cascade might have an additional strong effect on the turbulent dissipation rate of dominantly outward propagating Alfv\'enic waves in structured plasma, as in the solar corona and solar wind.

\end{abstract}


\section{Introduction} \label{one}

Stemming from the remarkable analogy of the magnetohydrodynamic (MHD) equations to the hydrodynamic equations when expressed in the formalism now bearing his name, \citet{1950PhRv...79..183E} was probably the first to point out that \textit{`one might,
in particular, expect that phenomena of turbulence will occur in
hydromagnetic systems (...). They will no doubt give rise to a ``turbulent" magnetic field coupled with the mechanical motion.'} Since then, this possibility was confirmed by numerous direct and indirect observations, both for astrophysical and laboratory plasmas \citep[see, e.g.][]{1995SSRv...73....1T,2013LRSP...10....2B,2014PSST...23f3001B}. One of the first theoretical breakthroughs towards understanding MHD turbulence was put forth by \citet{1964SvA.....7..566I} and \citet{1965PhFl....8.1385K}. They pointed out that the nonlinear, turbulent cascade of energy is the result of collisions (mutual deformation) of oppositely propagating Alfv\'enic wave-packets. This result was based on the assumption of an incompressible and homogeneous plasma. Under these conditions, arbitrarily nonlinear pure Alfv\'en waves are exact solutions to the MHD equations. In the Els\"{a}sser formalism, a pure Alfv\'en wave is completely described by one of the Els\"{a}sser variables, $\textbf{z}^\pm = \textbf{v} \pm \textbf{B}/\sqrt{ \mu_0 \rho}$, with the sign representing parallel ($-$) or anti-parallel ($+$) propagation, with respect to the mean magnetic field. Nonlinear interactions occur only when both $\textbf{z}^\pm$ are nonzero, thus the need for counterpropagating waves. This central idea is still generally accepted as the basis of MHD turbulence \citep[e.g.,][]{2013PhPl...20g2302H}. For a broader perspective on MHD turbulence, see the review in \citet{2004RvMP...76.1015Z}. Indeed, it is often not realized that once the assumptions of incompressibility or homogeneity, or both, do not apply, the MHD spectrum is much richer and allows for more complex dynamics than the propagation of pure Alfv\'en waves \citep{2011SSRv..158..289G,10.3389/fspas.2019.00020}; Compressibility allows for the existence of magnetoacoustic waves, which in a homogeneous plasma are the fast and slow waves. Inhomogeneities generally allow for the existence of surface and global waves, which have distinct properties compared to the normal modes of a homogeneous plasma \citep{2014masu.book.....P}. MHD waves in an inhomogeneous plasma are also referred to as waves with mixed properties, i.e. both Alfv\'en and magnetoacoustic properties. The mixed properties arise because in an inhomogeneous plasma the Eulerian perturbation of total pressure couples with the dynamics of the motion \citep{1982alwa.book.....H}. The linear coupling of the Els\"{a}sser variables due to density inhomogeneity (e.g. gravitational stratification in a coronal hole) along the mean magnetic field was shown by \citet{1980JGR....85.1311H}, for an incompressible plasma. This leads to Alfv\'en wave reflections. The inhomogeneity along the propagation direction acts as a source of inward-propagating Alfv\'en waves, even if only outward (away from the Sun) propagating waves are present initially. Then, the outgoing and reflected, counterpropagating Alfv\'en waves lead to turbulence due to the same nonlinear term involving both $\textbf{z}^\pm$ nonzero. Turbulence due to reflected waves is still an intense research topic getting considerable attention \citep{1999ApJ...523L..93M,2001PhPl....8.2377D,2003NPGeo..10...93M,2013ApJ...776..124P,
2014ApJ...782...81V,2016ApJ...821..106V}. \par
However,  \citet{1989PhRvL..63.1807V} and \citet{1990JGR....9514873H} pointed out that linear coupling also implies that modes present both $\textbf{z}^\pm$ nonzero while propagating. The secondary component of $\textbf{z}^-$ (say), co-propagating with $\textbf{z}^+$, was referred to as the `anomalous' component. \citet{1990JGR....9514873H} went a step further and stated, based on harmonic analysis, that an Alfv\'en wave propagating in a plasma which is inhomogeneous along the magnetic field is necessarily described by both $\textbf{z}^+ \neq 0$ and $\textbf{z}^- \neq 0$, and therefore it is `\textit{incorrect to make the (common) assumption that an observation of $\textbf{z}^-$ (say) represents an outward propagating Alfv\'en wave while $\textbf{z}^+$ necessarily represents an inward propagating Alfv\'en wave}'.  \citet{1989PhRvL..63.1807V} proposed that the `anomalous' $\textbf{z}^-$ fields are co-propagating with the principal $\textbf{z}^+$ and can lead to a turbulent cascade which is essentially different from that in homogeneous MHD turbulence due to counterpropagating waves. They argued that, as $\textbf{z}^\pm$ are propagating in the same direction with the same velocity, the nonlinear interaction is coherent, modifying the spectral characteristics of the resulting turbulence. The idea of nonlinearly interacting co-propagating $\textbf{z}^\pm$ disturbances was met initially with opposition \citep{1990PhRvL..64.2591Z,1990PhRvL..64.2592V}. Nevertheless, the idea that there are co-propagating $\textbf{z}^\pm$ fields was accepted for nearly two decades, until \citet{2007JGRA..112.8102H} pointed out that the harmonic analysis in \citet{1990JGR....9514873H} led to wrong conclusions. Using an impulse function approach, \citet{2007JGRA..112.8102H} show that while the $\textbf{z}^\pm$ fields are indeed coupled for an outward-propagating Alfv\'en wave, this does not mean that the `anomalous' Els\"{a}sser field is co-propagating with the principal one. The leading edge of the `anomalous' component may seem to propagate outward, however this does not imply
that it follows anything other than the inward characteristics
in the plasma frame. \citet{2007JGRA..112.8102H} explain: `\textit{A simple analogy might be smoke coming out of the smokestack of a ship that is steaming into a headwind. The front of the smoke trail is always at the forward moving smokestack, but each smoke particle is blown backward by the wind as soon as it leaves the smokestack}'. Although this study dismisses the existence of an `anomalous' field which is co-propagating \citep[for the specific setup as in][]{1980JGR....85.1311H}, it does not change the fact that there can be coherent nonlinear interactions between $\textbf{z}^\pm$, e.g. at the leading edges of wave packets. What changes is that technically these can still be categorized in the oppositely propagating wave interaction description, and do not represent self-deformations of waves. The implications of such coherent nonlinear interactions were since then exploited in a number of studies \citep{2009ApJ...700L..39V,2012ApJ...750L..33V}. As we will see later, a plasma with density variations (stratification) along the field but otherwise permeated by a straight and uniform magnetic field is a very special case. Under these conditions, one can still globally talk about linearly coupled outward and inward-propagating pure Alfv\'en waves. Therefore, it is also a special inhomogeneous setup in which the Els\"{a}sser variables still retain their identity, i.e. $\textbf{z}^+$ and $\textbf{z}^-$ still represent strictly in or outward-propagating Alfv\'en waves. This peculiarity might be the reason why the existence of truly co-propagating Els\"{a}sser fields was missed. In general, the presence of co-propagating Els\"{a}sser fields just means that the waves are no longer pure Alfv\'en waves encountered in a homogeneous plasma. For example, magnetosonic waves in a homogeneous plasma generally present both Els\"{a}sser variables while propagating, as also do all waves in a generally inhomogeneous plasma \citep{2019ApJ...873...56M}. Therefore, in this study, we shall avoid calling the induced co-propagating Els\"{a}sser component `anomalous', as it was customary in the literature. Indeed, there is nothing anomalous in an MHD wave that is not a pure Alfv\'en wave due to the presence of inhomogeneities or compression. \par
Turbulence induced by truly unidirectionally-propagating waves proposed in \citet{2017NatSR...14820M}, which we refer to as `uniturbulence' or self-cascade of waves, is still largely unknown and it is controversial. This lingering disbelief (since the proposal by \citet{1989PhRvL..63.1807V}) might come from the erroneous conclusion that nonlinear interactions necessarily imply counterpropagating waves, as results from the incompressible and homogeneous MHD equations, or as might result from the setups involving Alfv\'en wave reflection, described above. \citet{1989JPlPh..41..479M} and \citet{1989GeoRL..16..755Z} derived the incompressible MHD equations in the Els\"{a}sser formalism for a generally inhomogeneous background plasma. These equations show clearly that inhomogeneities act as sources for the $\textbf{z}^\pm$ fields, through which they are linearly coupled. However, this does not necessarily mean only reflection (i.e. creation of counterpropagating waves). It also means that unidirectionally propagating waves are now described by both $\textbf{z}^\pm$ nonzero. Furthermore, in the presence of density inhomogeneities, there are nonzero nonlinear terms involving only either $\textbf{z}^+$ or $\textbf{z}^-$, a deviation from the original \citet{1965PhFl....8.1385K} picture. Departure from this picture is also present in homogeneous compressible turbulence, for which the exact relation includes multiple terms without the need for both $\textbf{z}^\pm$ nonzero \citep{2013PhRvE..87a3019B}. While inhomogeneities along the magnetic field are a popular initial condition in MHD turbulence studies, there are much fewer studies which deal with inhomogeneities across the magnetic field. \citet{1992ApJ...396..297M} simulated a 2D incompressible plasma with a straight but transversely varying magnetic field. They showed that the initial perturbation consisting of only $\textbf{z}^+$ quickly leads to the generation of $\textbf{z}^-$ through the source terms mentioned above. The wave is then phase-mixed, \citep[a linear process,][]{1983A&A...117..220H} in the direction of the inhomogeneity, leading to increasingly oblique wavefronts in time. The authors show (through energy spectra) that nonlinear interactions do occur between $\textbf{z}^\pm$, but as we will show later, they apparently miss the fact that the generated $\textbf{z}^-$ is co-propagating with $\textbf{z}^+$ and attributed it to `\textit{waves propagating in opposite directions}'. \citet{1998JGR...10323691G} studied a setup with pressure-balanced structures, and also showed the generation of $\textbf{z}^-$ (say) from a spectrum of pure $\textbf{z}^+$. However, they do not specify the propagation direction of $\textbf{z}^-$, focusing mostly on `refraction of parallel-propagating Alfv\'en waves to oblique angles', which is essentially the process of phase mixing. \par 
In this paper, we explore plasmas that are inhomogeneous perpendicularly to a straight magnetic field, both analytically and through numerical simulations. We show that unidirectionally propagating waves lead to nonlinear cascade of energy to higher perpendicular wavenumbers, i.e. they self-cascade. This paper is intended to offer a deeper insight into the phenomenon of uniturbulence, and to further bring this new turbulence generation mechanism to the attention of the MHD turbulence community. In the following, in Section~\ref{two} we start from the incompressible MHD equations to show the wave properties of perpendicularly inhomogeneous plasmas and the peculiarity of the longitudinally inhomogeneous case. In Section~\ref{three}, we show the results of 3D MHD simulations and the self-cascade of wave energy in a simple inhomogeneous setup. Finally, in Section~\ref{four}, we conclude the presented results.

\section{Inhomogeneous incompressible MHD} \label{two}

We start from the ideal, incompressible MHD equations \citep[e.g.,][]{2004prma.book.....G}:
\begin{align}
 \label{continuity}
 &\frac{\partial \rho}{\partial t} = -\mathbf{v} \cdot \nabla \rho, \\
 \label{motion}
 &\rho \frac{\partial \mathbf{v}}{\partial t} + \rho \mathbf{v} \cdot \nabla \mathbf{v} = -\nabla p + \mathbf{j} \times \mathbf{B}, \\
 \label{induction}
 &\frac{\partial \mathbf{B}}{\partial t} = \nabla \times (\mathbf{v} \times \mathbf{B}), \\
 \label{divb}
 &\nabla \cdot \mathbf{B} = 0,
\end{align}
where $\mathbf{j} = \frac{1}{\mu}(\nabla \times \mathbf{B})$ is the current density. Note that the solenoidal condition on the velocity, i.e. incompressibility ($\nabla \cdot \mathbf{v} = 0$) is implied by the formulation of the continuity equation in Eq.~\ref{continuity}. We retain this equation as we allow for density inhomogeneities. 
Using the Els\"{a}sser variables \citep{1950PhRv...79..183E}, defined as $\textbf{z}^\pm = \mathbf{v} \pm \mathbf{v}_A$, where $\mathbf{v}_A = \mathbf{B}/\sqrt{\mu \rho}$, the system of incompressible ideal MHD equations can be transformed into a simpler system  \citep{1987JGR....92.7363M,1989JPlPh..41..479M,1989GeoRL..16..755Z}:

\begin{equation}
\frac{\partial \mathbf{z}^\pm}{\partial t} + \mathbf{z}^\mp \cdot \nabla \mathbf{z}^\pm = - \frac{1}{\rho} \nabla P - \mathbf{v}_A ( \nabla \cdot \mathbf{v}_A),
\label{Elsasser}
\end{equation}
where $P = p + \frac{B^2}{2\mu}$ is the total pressure. The last term on the RHS can also be expressed using the Els\"{a}sser variables, as $\mathbf{v}_A = \frac{1}{2}(\mathbf{z}^+ - \mathbf{z}^-)$. These equations allow for general initial conditions, i.e. including inhomogeneous density, magnetic field and velocity, and arbitrary amplitudes, i.e. they represent the full nonlinear equations. The total pressure satisfies a Poisson equation, by taking the divergence of Eq.~\ref{Elsasser}:
\begin{equation}
\nabla ^2 P = - \nabla \cdot \left [ \rho \left( \frac{\partial \mathbf{z}^\pm}{\partial t} + \mathbf{z}^\mp \cdot \nabla \mathbf{z}^\pm + \mathbf{v}_A ( \nabla \cdot \mathbf{v}_A) \right) \right]
\label{Poisson_comp}
\end{equation}
In the absence of initial density inhomogeneities, $\rho(t) = \rho_0 = \mathrm{const}.$ as per Eq.~\ref{continuity}, and the divergence of $\textbf{z}^\pm$ and $\mathbf{v}_A$ vanishes. In this case the last term on the RHS of Eq.~\ref{Elsasser} vanishes, and the total pressure perturbation satisfies:
\begin{equation}
\nabla ^2 P = - \rho_0 \nabla \cdot (\mathbf{z}^\mp \cdot \nabla \mathbf{z}^\pm) = - \rho_0 \sum_i \sum_j \frac{\partial}{\partial i} z_j^- \frac{\partial}{\partial j} z_i^+.
\label{Poisson}
\end{equation}
\paragraph{The non-equivalence of gradients in magnetic field and density.}
Note that the finite amplitude equations in the absence of density inhomogeneities, albeit with an inhomogeneous magnetic field, are the same as in the case of a homogeneous and incompressible plasma:
\begin{equation}
\frac{\partial \mathbf{z}^\pm}{\partial t} + \mathbf{z}^\mp \cdot \nabla \mathbf{z}^\pm = - \frac{1}{\rho} \nabla P.
\label{Elsasser_homog}
\end{equation}
 Therefore density inhomogeneities appear to play a special role. At a first glance this might appear as a paradoxical outcome, as one might think that waves should be influenced not by density or magnetic field gradients, but by propagation speed gradients, here the Alfv\'en speed $v_A = B/\sqrt{\mu \rho}$. Nevertheless, by investigating Eq.~\ref{Elsasser}, it appears that in incompressible MHD one cannot derive a general inhomogeneous equation in which the coefficients are expressed only through the Alfv\'en speed. One might also think that it is certainly possible to have the same Alfv\'en speed gradient as a result of either density or magnetic field variations. However, this is not always true. This shows the peculiarity of the case discussed in the Introduction, that of a density-stratified plasma along a straight and uniform magnetic field. This initial condition cannot be achieved by magnetic field variations alone: a gradient in the magnetic field intensity along its direction would necessarily imply variations in perpendicular directions, to satisfy the solenoidal condition (although one might argue that if the solenoidity is satisfied in 2D, pure Alfv\'en waves could still exist in the third direction, however this leads to a different evolution of the waves). This observation hints at why the last term on the RHS of Eq.~\ref{Elsasser} is peculiar, and is only present for inhomogeneous density conditions. 
\subsection{A plasma with density inhomogeneities along the magnetic field}

Let us further investigate the peculiarity of the case presented above, as it can provide deep insights and comparison for the perpendicularly inhomogeneous cases. First, let us separate the background values and Eulerian perturbations, by writing the Els\"{a}sser variables as $\textbf{z}^\pm = \mathbf{z}_0^\pm + \delta \mathbf{z}^\pm$. We then take $\mathbf{z}_0^\pm = \pm B_0 \mathbf{\hat{x}}/\sqrt{\mu \rho_0(x)} = v_{A0}(x) \mathbf{\hat{x}}$, where $\mathbf{\hat{x}}$ is the direction of the homogeneous magnetic field. In the following, we take $\mu = 1$ and will drop the $\delta$ for perturbations for brevity. The linearized Eq.~\ref{Elsasser} then reads:
\begin{equation}
\frac{\partial \mathbf{z}^\pm}{\partial t}  \mp v_{A0}(x)\frac{\partial \mathbf{z}^\pm}{\partial x}  \pm \mathbf{z}^\mp_\parallel \frac{\partial v_{A0}(x)}{\partial x}\mathbf{\hat{x}} = - \frac{1}{\rho_0(x)} \nabla P - \frac{1}{2}(\mathbf{z}^+ - \mathbf{z}^-)\frac{\partial v_{A0}(x)}{\partial x} - \frac{1}{2} v_{A0}(x) \frac{\partial (\mathbf{z}_\parallel^+ - \mathbf{z}_\parallel^-)}{\partial x}\mathbf{\hat{x}},
\end{equation} 
where the subscript $\parallel$ stands for the component parallel to $\mathbf{\hat{x}}$. The parallel components are not coupled linearly to the perpendicular components: if we consider a purely outgoing Alfv\'en wave initially, with say $\delta \mathbf{z}^- \neq 0, \delta \mathbf{z}^+ = 0$, then the parallel components $\delta \mathbf{z}^\pm_\parallel$ remain zero, i.e. they have no source terms. Here we shall note that the parallel component in this case would be associated with pseudo-Alfv\'en waves, the incompressible vestige of slow waves, which share the dispersion relation with pure Alfv\'en waves, but have perpendicular polarization to these and components parallel to the magnetic field \citep[e.g.,][]{2003ASSL..294.....G}. We will not focus on pseudo-Alfv\'en waves here and consider just Alfv\'en waves. Instead, let us investigate the total pressure term, by decomposing it into the gas and magnetic pressure contributions:
\begin{equation}
\nabla P = \nabla p + \nabla \frac{B^2}{2} = \nabla p + \nabla \frac{1}{2}\rho v_A^2.
\end{equation}
The linear contribution of $v_A^2$ is $(\mathbf{v}_{A0}+\delta\mathbf{v}_A) \cdot (\mathbf{v}_{A0}+\delta\mathbf{v}_A) = 2 v_{A0}(\delta \mathbf{z}_\parallel^+ - \delta \mathbf{z}_\parallel^-)$. As we can see, the magnetic pressure term is nonzero linearly only when the parallel components of the Els\"{a}sser variables are not vanishing. This will have crucial implications later on. However, for this case, it is just a confirmation of the known fact that pure Alfv\'en waves present no linear magnetic pressure perturbations. As pure Alfv\'en waves present neither gas pressure perturbations linearly, the total pressure term is zero, meaning that the dynamics are described by setting the argument of the divergence in Eq.~\ref{Poisson_comp} to zero. The linear version of this equation is analogous to the equations derived by \citet{1980JGR....85.1311H}:
\begin{equation}
\frac{\partial \mathbf{z}^\pm_\perp}{\partial t}  \mp v_{A0}(x)\frac{\partial \mathbf{z}^\pm_\perp}{\partial x}  = - \frac{1}{2}(\mathbf{z}_\perp^+ - \mathbf{z}_\perp^-)\frac{\partial v_{A0}(x)}{\partial x},
\end{equation}
where the subscript $\perp$ stands for the perpendicular component. These equations describe outward and inward-propagating pure Alfv\'en waves which are linearly coupled due to reflections and completely described by $\mathbf{z}^\pm_\perp$: they are advected in opposite directions \citep{2007JGRA..112.8102H}. As we will see shortly, the reason why pure Alfv\'en waves still exist in this case is the lack of perpendicular inhomogeneities, in the perturbation direction. \par

\subsection{A plasma with inhomogeneities perpendicular to the magnetic field}

Let us now consider the case of perpendicular structuring. Although density or magnetic field variations in the perpendicular direction are different, in the case of only perpendicular structuring the last term on the RHS of Eq.~\ref{Elsasser} has linear components only along the field, and as we will see this component does not `feed' the perpendicular component. For similar Alfv\'en speed gradients, the evolution should not be substantially different. Therefore, we consider magnetic field variations as then the last term on RHS of Eq.~\ref{Elsasser} vanishes. We will consider a straight magnetic field which is varying in the $\mathbf{\hat{z}}$ direction: $\mathbf{z}_0^\pm = \pm B_0(z) \mathbf{\hat{x}}/\sqrt{\rho_0} = \pm v_{A0}(z)\mathbf{\hat{x}}$. The linearized Els\"{a}sser equations now take the form: 
\begin{equation}
\frac{\partial \mathbf{z}^\pm}{\partial t}  \mp v_{A0}(z)\frac{\partial \mathbf{z}^\pm}{\partial x}  \pm \mathbf{z}^\mp_z \frac{\partial v_{A0}(z)}{\partial z}\mathbf{\hat{x}} = - \frac{1}{\rho_0} \nabla P,
\label{Elsasser_inhomperp}
\end{equation} 
where the $z$ subscript stands for the $\mathbf{\hat{z}}$ component of $\mathbf{z}^\pm$. First we consider a perturbation in the $\mathbf{\hat{y}}$ component of say, $\mathbf{z}^+$. With only $\mathbf{z}^+_y$ perturbed (uniformly along the $y$-direction), this component is decoupled from the other components, and it travels as a pure Alfv\'en wave, with different speed on different field lines:
\begin{equation}
\frac{\partial \mathbf{z}_y^\pm}{\partial t}  = \pm v_{A0}(z)\frac{\partial \mathbf{z}_y^\pm}{\partial x}.
\end{equation} 
This is the phenomenon of phase mixing \citep{1983A&A...117..220H}: Alfv\'en waves traveling on neighbouring field lines will get out of phase if there is an Alfv\'en speed gradient across the field lines, leading to a bending of the wavefronts, or in the formulation of e.g. \citet{1998JGR...10323691G}, a refraction of the Alfv\'en waves to oblique angles. There is no coupling between $\mathbf{z}^\pm$, and the Els\"{a}sser variables completely describe the phase mixed Alfv\'en waves traveling up (say, $\mathbf{z}^-$) or down ($\mathbf{z}^+$) the magnetic field lines. As we will see in the following, the $\mathbf{z}_y^\pm$ components are coupled linearly to $\mathbf{z}_z^\pm$ in this equilibrium if an initial perturbation in $\mathbf{z}_z^\pm$ has a $\mathbf{\hat{y}}$-dependence.  \par 
Now we will focus on the main subject of this paper, that of inhomogeneities in the wave perturbation direction, perpendicular to the magnetic field, and we will indicate how it leads to a different turbulence phenomenology, that of self-cascading waves. The study of MHD waves in plasmas in which there is an inhomogeneity in the direction of the perturbation has a long history \citep[e.g.,][]{1964AnPhy..29..282B,1971JPlPh...5..239S,1973ZPhy..261..217G}, and it is known to lead to the phenomena of MHD surface waves in the case of a discontinuous variation \citep[e.g.,][]{1964ApJ...139..690P,1974SoPh...37..127P,1981SoPh...69...27R,1992SoPh..138..233G,
2009A&A...503..213G}, or to quasi/global MHD waves and resonant absorption in the case of a continuous variation \citep[e.g.,][]{1974PhFl...17.1399C,1974PhRvL..32..454H,1991SoPh..133..227S}. A common and much-studied example of surface waves are the surface Alfv\'en waves in flux tubes \citep{2012ApJ...753..111G}, which are analogous to surface Alfv\'en waves on a planar interface in Cartesian geometry, as employed here \citep{1981SoPh...69...27R}. \par 
We consider initial perturbations to say, $\mathbf{z}^+$ in the $\mathbf{\hat{z}}$ direction, while $\mathbf{z}^- = 0$. We can immediately see a crucial difference between this case and the previously employed cases: now the parallel components of $\mathbf{z}^\pm$ are linearly coupled to the $\mathbf{\hat{z}}$ components through the inhomogeneity by the 3rd term in the LHS of Eq.~\ref{Elsasser_inhomperp}:
\begin{equation}
\frac{\partial \mathbf{z}_\parallel^\pm}{\partial t}  \mp v_{A0}(z)\frac{\partial \mathbf{z}_\parallel^\pm}{\partial x} = \mp \mathbf{z}^\mp_z \frac{\partial v_{A0}(z)}{\partial z}- \frac{1}{\rho_0} \nabla_\parallel P,
\label{Elsasser_par}
\end{equation}
where $\nabla_\parallel = \frac{\partial}{\partial x}\mathbf{\hat{x}}$.
Note that the perpendicular components act as source for the parallel components, but not the other way around:
\begin{equation}
\frac{\partial \mathbf{z}_\perp^\pm}{\partial t}  \mp v_{A0}(z)\frac{\partial \mathbf{z}_\perp^\pm}{\partial x} = -\frac{1}{\rho_0} \nabla_\perp P,
\label{Elsasser_perp}
\end{equation}
where $\nabla_\perp = \frac{\partial}{\partial y}\mathbf{\hat{y}} + \frac{\partial}{\partial z}\mathbf{\hat{z}}$.
 This can be interpreted as a single wave perturbing both the parallel and perpendicular components. Therefore the resulting waves necessarily have both $\mathbf{z}^\pm$ nonzero, and also both parallel and perpendicular components. The other crucial difference is that the linear total pressure term is nonzero. We can immediately see this from Eq.~\ref{Poisson}, and our previous discussion on the magnetic pressure term. The resulting waves cause linear total pressure perturbations as they propagate. This is important, as harmonic modes of homogeneous and incompressible plasmas show no total pressure perturbations. Indeed, these waves can only exist due to the inhomogeneity, and can be referred to as surface Alfv\'en waves in case of a discontinuous variation or quasi/global Alfv\'en waves or Alfv\'enic waves in the case of continuous variation \citep[e.g.,][]{2002ESASP.505..137G}. It is the linear total pressure term which couples the perpendicular components of $\mathbf{z}^\pm$. Note that if the initial perturbation in $\mathbf{z}_z^+$ has a $\mathbf{\hat{y}}$-dependence, the linear total pressure term also couples the $\mathbf{z}_z^\pm$ and $\mathbf{z}_y^\pm$ components, as the $\mathbf{\hat{y}}$-dependence of $\mathbf{z}_z^+$ implies the $\mathbf{\hat{y}}$-dependence of the linear pressure term through Eq.~\ref{Poisson}. Therefore, in the following we analyze the perpendicular component $\mathbf{z}_\perp^\pm$ instead of the $\mathbf{\hat{z}}$ and $\mathbf{\hat{y}}$ components for a more general presentation, while understanding that $\mathbf{z}_z^\pm$ is nonzero. \par
It is well known that the surface Alfv\'en or quasi/global Alfv\'en waves are propagating wave solutions. That is, unidirectionally propagating while suffering no backward reflections. This hints that the Els\"{a}sser variables are coupled and co-propagating, as they both describe a wave propagating in one direction. As we mentioned in the Introduction, this just means that these waves are not pure Alfv\'en waves any longer, hence they must necessarily be described by both Els\"{a}sser variables while propagating \citep{2019ApJ...873...56M}. In fact, this can be shown, if a Fourier analysis of the perpendicular components is possible in time and in the $\mathbf{\hat{x}}$ direction:
\begin{equation}
\mathbf{v}_\perp = \hat{v}(y,z) e^{i(k x - \omega t)},\qquad \mathbf{B}_\perp = \hat{B}(y,z) e^{i(k x - \omega t)},
\label{Fourier}
\end{equation}
where for positive $k$ and $\omega$, the waves are propagating towards positive $x$-values in time. For the sake of a more general derivation, let us consider for now a generally inhomogeneous plasma in the perpendicular direction, i.e. $\rho_0(y,z), B_0(y,z)$. The perpendicular component of the induction equation (Eq.~\ref{induction}) reads:
\begin{equation}
\frac{\partial \mathbf{B}_\perp}{\partial t} = B_0(y,z) \frac{\partial \mathbf{v}_\perp}{\partial x},
\label{Indperp}
\end{equation}
which Fourier-analyzed according to Eqs.~\ref{Fourier} gives:
\begin{equation}
\hat{B}(y,z) = -B_0(y,z)\frac{k}{\omega}\hat{v}(y,z) = -\frac{B_0(y,z)}{v_{ph}}\hat{v}(y,z),
\end{equation}
where $v_{ph} = \omega/k$ is the phase speed determined by the dispersion relation of the specific initial condition. Using $\mathbf{v} = \frac{1}{2}(\mathbf{z}^+ + \mathbf{z}^-)$ and $\mathbf{B} = \frac{1}{2}\sqrt{\rho}(\mathbf{z}^+ - \mathbf{z}^-)$ from the definition of Els\"{a}sser variables, the Fourier-analyzed induction equation gives:
\begin{equation}
\left(v_{ph} + v_{A0}(y,z) \right) \mathbf{z}_\perp^+ = \left(v_{ph} - v_{A0}(y,z)\right)\mathbf{z}_\perp^-,
\label{CoupledElsasser}
\end{equation}
This is an important relation of the perpendicular components of the Els\"{a}sser variables. Note that when $v_{ph} = v_{A0} = \mathrm{const.}$, as in a homogeneous incompressible plasma the fields are not coupled. Eq.~\ref{CoupledElsasser} is a very general result. Specifically, it applies to both compressible and incompressible plasmas, as Eq.~\ref{Indperp} is generally valid. Thus it relates the perpendicular components of the Els\"{a}sser variables in the linear regime in any plasma configuration homogeneous along the magnetic field.\par 
Returning to a magnetic field inhomogeneity along $\mathbf{\hat{z}}$ as considered initially, let us further investigate Eqs.~\ref{Elsasser_perp}. They show that starting from only $\mathbf{z}_\perp^+$, the source of $\mathbf{z}_\perp^-$ is the linear total pressure perturbation. A naive interpretation of these equations would lead us to conclude that $\mathbf{z}_\perp^\pm$ are counter-propagating, since the speeds in front of the advective term are multiplied by $\mp 1$, in direct contradiction with the well-known fact that solutions to Eq.~\ref{Elsasser_inhomperp} are propagating waves which do not display wave reflection. Therefore, the total pressure term has to act in such a way as to cancel the backward advection of $\mathbf{z}_z^-$. However, this cannot be its only function, as both $\mathbf{z}_\perp^\pm$ should exist. It also acts as a source for $\mathbf{z}_\perp^-$. Indeed, if we replace the total pressure term in Eqs.~\ref{Elsasser_perp} for $\mathbf{z}_\perp^-$ from the equation for $\mathbf{z}_\perp^+$, we get:
\begin{equation}
\frac{\partial \mathbf{z}_\perp^-}{\partial t} - \frac{\partial \mathbf{z}_\perp^+}{\partial t} = -v_{A0}(z)\frac{\partial \mathbf{z}_\perp^-}{\partial x}-v_{A0}(z)\frac{\partial \mathbf{z}_\perp^+}{\partial x}.
\end{equation}
If we substitute either $\mathbf{z}_\perp^\pm$ by using Eq.~\ref{CoupledElsasser}, we get the important result:
\begin{equation}
\frac{\partial \mathbf{z}_\perp^\pm}{\partial t} = -v_{ph}\frac{\partial \mathbf{z}_\perp^\pm}{\partial x},
\label{Surface}
\end{equation}
which confirms our previous claim that the Els\"{a}sser variables both propagate in the same direction, with speed $v_{ph}$. Note that Eq.~\ref{Surface} is still valid for a general inhomogeneity across the magnetic field, $v_{A0}(y,z)$. For arbitrary $v_{A0}(y, z)$, $v_{ph}$ has no analytical solution. Numerically it can be found by using e.g. the T-Matrix theory \citep{1994ApJ...436..372K,2009ApJ...692.1582L,2010ApJ...716.1371L}. However, in the case of a discontinuous variation of $B_0(z)$, i.e. an interface at $z = 0$, the solutions to Eq.~\ref{Elsasser_inhomperp} are known \citep{1981SoPh...69...27R}, and the dispersion relation gives:
\begin{equation}
v_{ph} = \sqrt{\frac{\rho_i v_{Ai}^2+\rho_e v_{Ae}^2}{\rho_i+\rho_e}},
\end{equation}
where the $i$ and $e$ subscripts represent the values above and below $z=0$, respectively. Then Eq.~\ref{Surface} just describes surface Alfv\'en waves. \par
After the realization that a plasma with inhomogeneities perpendicular to the magnetic field admit as linear solutions unidirectionally propagating waves which are necessarily described by both $\mathbf{z}^\pm$ nonzero and co-propagating, let us get back to the equation describing the nonlinear evolution, now with $v_{A0}(y,z)\mathbf{\hat{x}}$:
\begin{equation}
\frac{\partial \mathbf{z}^\pm}{\partial t}  \mp v_{A0}(y,z)\frac{\partial \mathbf{z}^\pm}{\partial x}  \pm \mathbf{z}^\mp_\perp \nabla_\perp v_{A0}(y,z) \mathbf{\hat{x}} + \mathbf{z}^\mp \cdot \nabla \mathbf{z}^\pm = - \frac{1}{\rho_0} \nabla P,
\label{fullperp}
\end{equation} 
The nonlinear advective term (last term on LHS), responsible for turbulence generation in a homogeneous and incompressible plasma, is still the essential nonlinearity here, still requiring both $\mathbf{z}^\pm$ nonzero. However, now wave solutions are described by $\mathbf{z}^\pm$ which are co-propagating! This leads to a coherent interaction between $\mathbf{z}^\pm$, which can be interpreted as a self-cascade or self-deformation of unidirectionally propagating waves, a phenomenon which leads to what we call uniturbulence. The coherent interaction differentiates uniturbulence from the counterpropagating wave phenomenology, in which interactions are incoherent deformations of Alfv\'en waves. There are also other differences between the two turbulence generation mechanisms. The nonlinear deformation of colliding Alfv\'en waves is an inherently three-dimensional phenomenon, as the waves must collectively vary along both perpendicular directions \citep{2013PhPl...20g2302H}. This criteria is expressed as $\mathbf{k}^+_\perp \times \mathbf{k}^-_\perp \neq 0$, where $\mathbf{k}^\pm_\perp$ are the perpendicular wave vectors. In uniturbulence, this is still a valid criteria for the existence of a self-cascade along both perpendicular directions, where one or both of the wave vectors can be given by the plasma inhomogeneity. However, the self-deformation of waves is no longer necessarily three-dimensional. This fact can be seen easily from the properties of surface/global waves arising in inhomogeneous plasmas. These waves generally have a varying perturbation amplitude in the perturbation direction. For example, a linearly polarized (two-dimensional) surface Alfv\'en wave has perpendicular velocity perturbations which decay exponentially away from the interface, resulting in perpendicular gradients of the Els\"{a}sser variables. As surface Alfv\'en waves have both $\mathbf{z}^\pm_\perp \neq 0$, the nonlinear advective term is nonzero. In fact, one-dimensional acoustic waves are well-known to self-deform, a process mostly referred to as nonlinear wave steepening or shock formation. The resulting turbulent dynamics is often referred to as burgulence \citep{2001ntt..conf..341F}.

\section{Numerical simulations of uniturbulence}\label{three}

The first numerical demonstration of the phenomenon of self-cascading, unidirectionally propagating waves was presented by \citet{2017NatSR...14820M}. They considered a setup with multiple random Gaussian density enhancements across the straight magnetic field, while along the magnetic field the plasma was homogeneous. This setup was too complicated to be easily analyzed and understood in simple terms regarding the generation of energy cascade. Here we shall present a simple workable example in which uniturbulence develops. 
\subsection{Numerical code, initial and boundary conditions}
We run full 3D ideal MHD simulations using the \texttt{FLASH} code \citep{2000ApJS..131..273F,2009ASPC..406..243L}, opting for the
third-order unsplit staggered mesh Godunov method \citep{2009JCoPh.228..952L,2013JCoPh.243..269L} with HLLD solver and \texttt{mc} slope limiter. The code implements constrained transport to keep the
divergence of the magnetic field down to round-off errors. An adaptively refined mesh is used, with three levels of refinement for the highest resolution runs. The base resolution is $64 \times 80 \times 120$ for a domain of size (in user units) of $0.6 \times 0.1 \times 0.15$. We found that higher resolution runs show generation of increasingly smaller scales and more complicated flow behaviour, as expected. However, this does not alter our qualitative conclusions. The origin lies in the geometric center of the bottom $\mathbf{\hat{y}-\hat{z}}$ slice. As we expect smooth dynamics along the magnetic field, we set a much lower resolution in the $\mathbf{\hat{x}}$ direction. We chose to present the results in user units, as the MHD equations are scale-invariant. In these units, the magnetic permeability $\mu$ is unity. Next, we will present the simple initial conditions of our model. We employ a straight, homogeneous magnetic field $B_0 \mathbf{\hat{x}}$, with $B_0 = 1.115$. This specific value has no special significance besides normalization purposes. The gas pressure is constant, as required for equilibrium, with the ratio of gas to magnetic pressure $\beta \approx 7.7$. As we solve the full compressible MHD equations, we opted for a high plasma beta in order to minimize the compressible contribution, i.e. as close as possible to the incompressible limit. Our discussion in Section~\ref{two} is valid for an incompressible plasma, however by setting our velocity perturbations much lower than both the Alfv\'en and sound speeds, we assure that compressible effects are kept at a minimum, i.e. the resulting waves are highly Alfv\'enic and presenting minimal compressive contributions. Nevertheless, tests with different plasma beta values (from $\beta \approx 0.02\ \mathrm{to}\ 15$) indicate that the dynamics perpendicularly to the magnetic field are not very sensitive to its value. The density is varying discontinuously at $z=0$ from 0.5 to 2.5, while it is constant along the other directions. Although there is a discontinuity at $t=0$ in density, this discontinuity is transitioning into a steep gradient once the simulation starts, as we have a nonzero numerical diffusion, inherent to the scheme. The thickness of this inhomogeneous layer is therefore determined by the resolution. We opted for a strong inhomogeneity as then the nonlinear terms are significantly large (in comparison to linear terms) even for relatively small Mach numbers of the perturbations. Moreover, we opted for an inhomogeneous density and not an inhomogeneous magnetic field as in Section~\ref{two} because jumps in magnetic field values are much less stable numerically. As discussed in Section~\ref{two}, this should not significantly affect the linear perpendicular evolution of the wave, however it adds an extra nonlinear term. We do not consider any equilibrium flows. \par
The boundary conditions are the Neumann-type zero-gradient or open conditions in the $\mathbf{\hat{z}}$ direction for all variables, and periodic boundaries in the $\mathbf{\hat{y}}$ direction. In the $\mathbf{\hat{x}}$ direction, at the bottom we impose a sinusoidal velocity driver polarized in the $\mathbf{\hat{z}}$ direction which is varying sinusoidally along the $\mathbf{\hat{y}}$ direction:
\begin{equation}
\mathbf{v}_z(y,t) = A\ \mathrm{cos}(\omega t)\ \mathrm{sin}(k_y y).
\end{equation}.
We set the frequency $\omega = 2\pi$ and the wavenumber $k_y = 2 \pi/0.1$, which corresponds to one wavelength fitting in the $\mathbf{\hat{y}}$ direction. Without $\mathbf{\hat{y}}$-variation in the driver, the problem would be two-dimensional, as the $\mathbf{\hat{y}}$ components are then decoupled, and we would not expect a cascade in both perpendicular directions, as stated in Section~\ref{two}. We run simulations with 2 amplitudes: a `low' amplitude ($A \approx 0.001$) and a `high' amplitude ($A \approx 0.01$) driver. The minimum Alfv\'en speed in the simulation is $V_{A0} \approx 0.7$, while the minimum sound speed is $c_{s0} \approx 1.26$. This shows that the Alfv\'en and sound Mach numbers are low, 0.014 and 0.008, respectively. At the top boundary in the $\mathbf{\hat{x}}$ direction we impose open boundary conditions, in order to allow waves to leave the domain freely. Tests with homogeneous density runs show maximum $1.0\%$ reflection of the incident Alfv\'en wave energy. 

\subsection{Simulation results}

\begin{figure}[h!]
  \centering
  \medskip
  \includegraphics[width=1.0\textwidth]{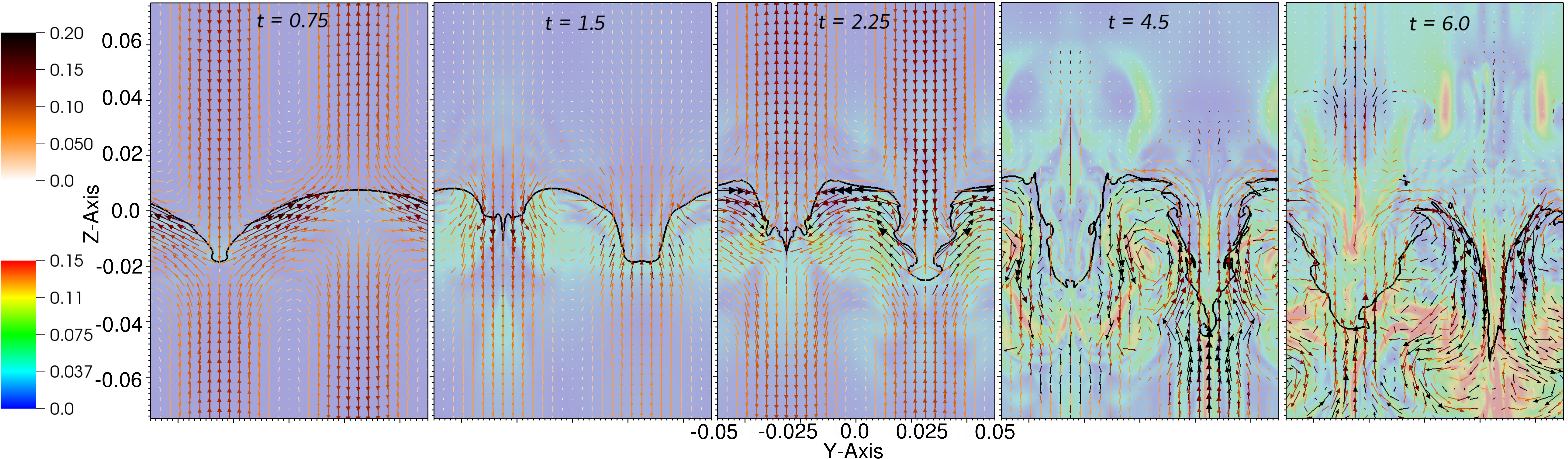}
  \caption{Snapshots of $\mathbf{z}_\perp^-$ (vector arrow field), and $|\mathbf{z}^+|$ (color plot) in the cross-section of the simulation box, at $x=0.5$, for the high amplitude run, shown at 5 different times, indicated on the top of each snapshot. The black contour is showing the deformation of the interface in density. Vector magnitude, besides by length, is also represented by the `orangehot' color table (on top). Vector origin is the middle of the arrows. Values are in user units.} 
  \label{Els_hi}
\end{figure}

\begin{figure}[h!]
  \centering
  \medskip
  \includegraphics[width=1.0\textwidth]{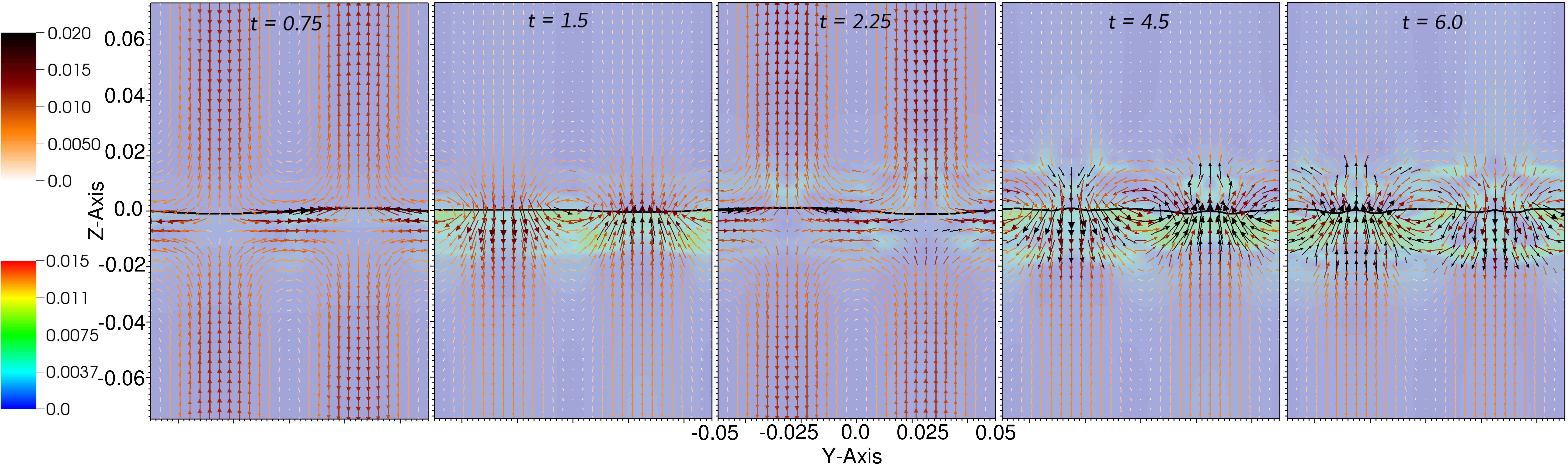}
  \caption{Same as in Figure~\ref{Els_hi}, but for the low amplitude run.} 
  \label{Els_lo}
\end{figure}

\begin{figure}[h!]
  \centering
  \medskip
  \includegraphics[width=0.75\textwidth]{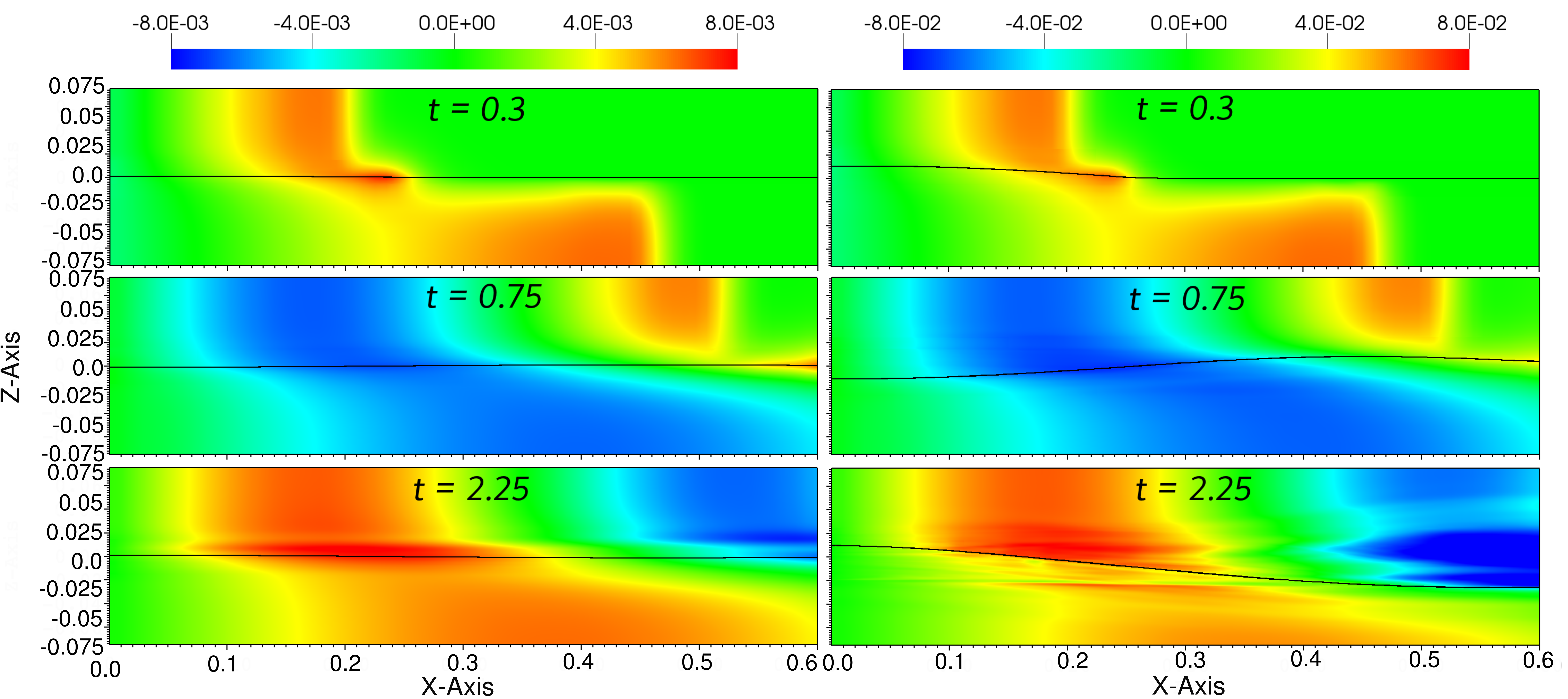}
  \caption{Snapshots of $v_z$ in the $x-z$ plane, at $y=0.025$, for the low (left column) and high (right column) amplitude runs, shown at 3 different times, indicated on the top of each snapshot. The black contour shows the position of the interface in density. Values are in user units.} 
  \label{long_cross}
\end{figure}

\begin{figure}[h!]
  \centering
  \medskip
  \includegraphics[width=1.0\textwidth]{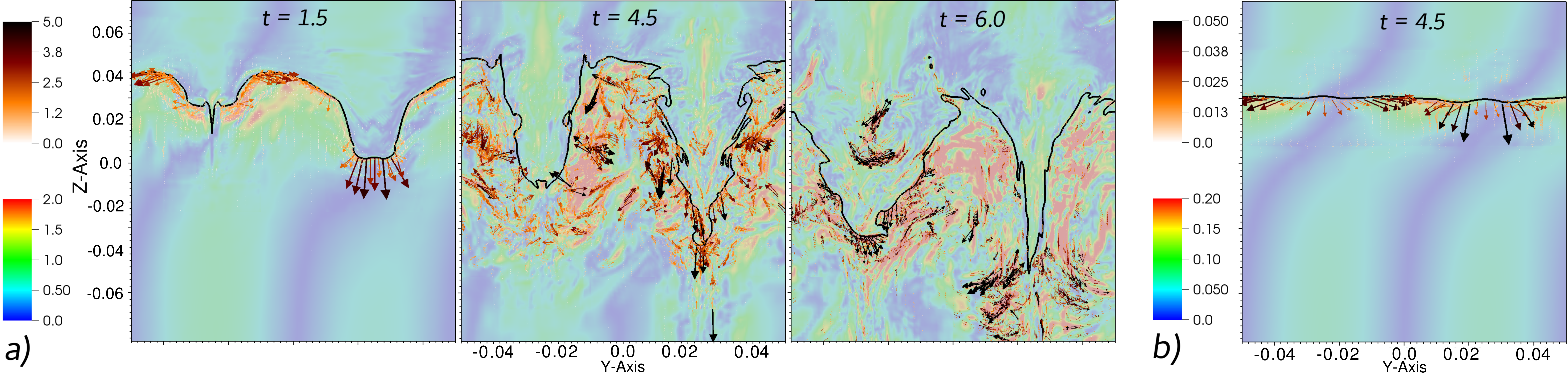}
  \caption{Snapshots of the nonlinear advection plus nonlinear density gradient terms \\ ($-\mathbf{z}^+ \cdot \nabla \mathbf{z}^- - \mathbf{v}_A ( \nabla \cdot \mathbf{v}_A)$, vector arrow field), and linear advection term ($v_{A0}(y,z)\partial_x |\mathbf{z}_\perp^-|$, color plot), in the cross-section of the simulation box, at $x=0.5$, shown at 3 different times, indicated on the top of each snapshot for the high amplitude run (left), and at $t=4.5$ for the low amplitude run (right). Vector magnitude, besides by length, is also represented by the `orangehot' color table (on top). Vector origin is the tail of the arrows, to better emphasize the location of nonlinearities. Values are in user units.} 
  \label{nonlin-advec}
\end{figure}

We run the simulation for 6 periods of the driver, that is, until $t=6.0$. In Figure~\ref{densplot} and Figure~\ref{densplot2}, the evolution of the density is shown in a cross-section for the high and low amplitude drivers, respectively. 
\begin{figure}[h!]
  \centering
  \medskip
  \includegraphics[width=1.0\textwidth]{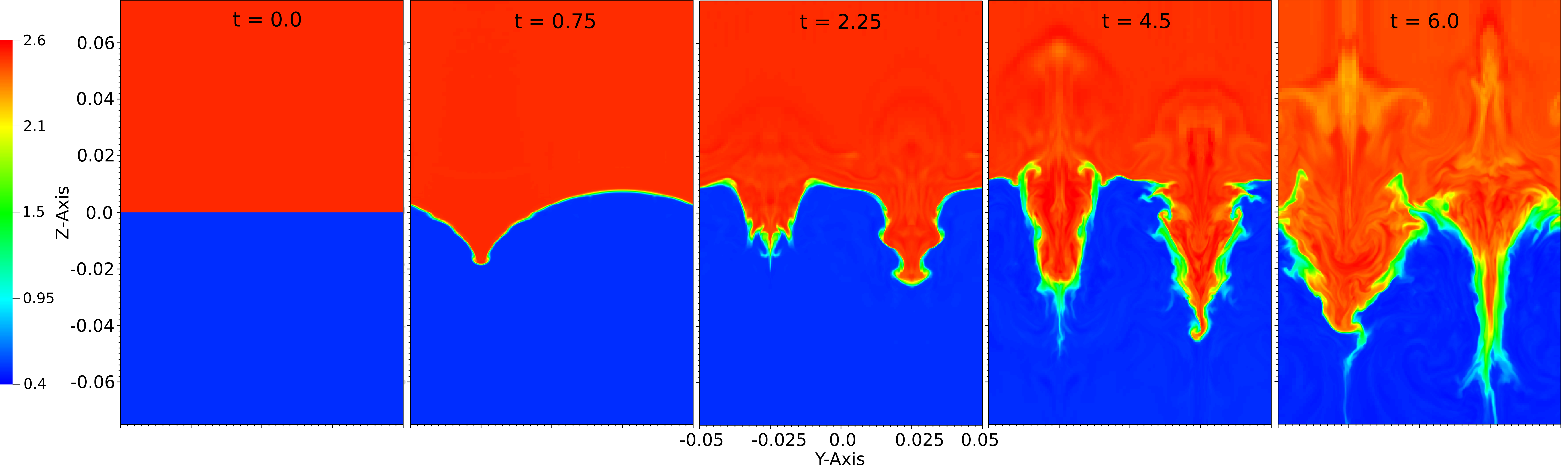}
  \caption{Snapshots of the density in the cross-section of the simulation box, at $x=0.5$, shown at 5 different times, indicated on the top of each snapshot, for the high amplitude run. The first snapshot shows the initial condition, a jump in density at $z=0$.} 
  \label{densplot}
\end{figure}
\begin{figure}[h!]
  \centering
  \medskip
  \includegraphics[width=1.0\textwidth]{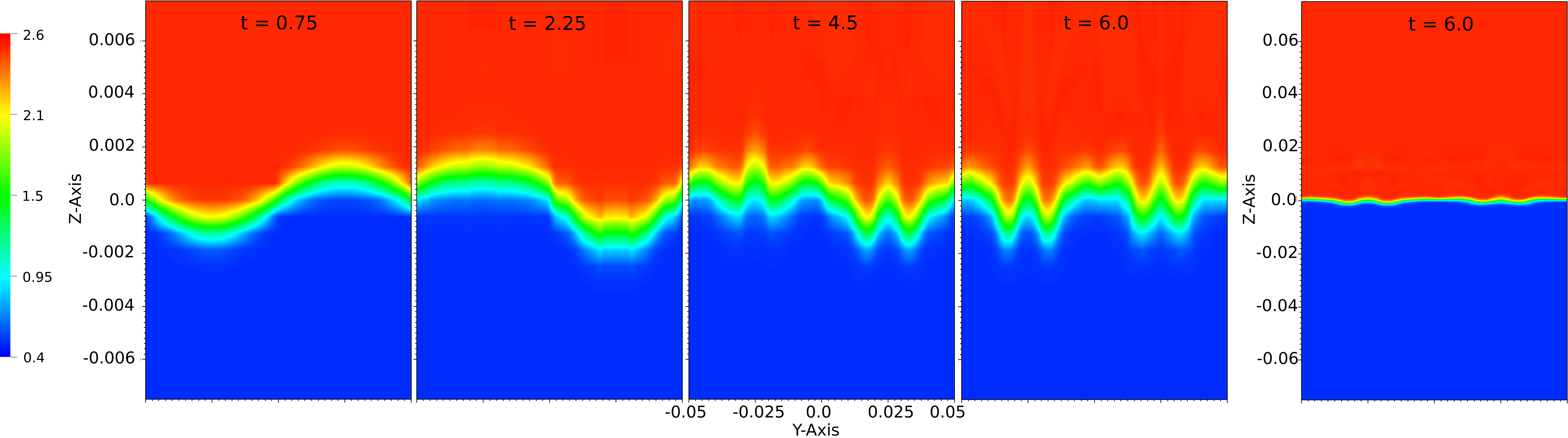}
  \caption{Snapshots of the density in the cross-section of the simulation box, at $x=0.5$, shown at 4 different times, indicated on the top of each snapshot, for the low amplitude run.  Note that for the sequence of snapshots on the left, the $z$-axis was magnified tenfold. The snapshot at $t=0.6$ on the right is not magnified, for comparison. The images were smoothed in order to avoid pixelation due to the limited numerical resolution. This might have introduced some artificial features.} 
  \label{densplot}
\end{figure}
The second snapshot from the left shows the first maximum displacement, at $t=0.75$. While the low amplitude solution at this time is almost a sinusoidal in the $y$-direction, in agreement with the linear solution \citep{1981SoPh...69...27R}, the high amplitude solution already shows deviations from the linear regime. There are `ripples' appearing on the density surface, as well as overall deformations, indicating that the nonlinear generation of smaller scales is already at work. At later times, we observe increasingly stronger and more complex deformations of the interface for the high amplitude run, but also deformations for the low amplitude run. Note that in the high amplitude run fluctuations do not only affect the interface, but are also visibly present in the homogeneous regions around the interface at later times. In Figure~\ref{Els_hi} and Figure~\ref{Els_lo}, the Els\"{a}sser fields are shown in the cross-section at different times, for the high and low amplitude drivers, respectively.
For waves of Alfv\'enic character propagating along the magnetic field pointing in the $x$ direction, the dominant Els\"{a}sser field is $\mathbf{z}^-$ (see Eq.~\ref{CoupledElsasser}), while $\mathbf{z}^+$ is relatively small. Note that while $\mathbf{z}^-$ is present in the whole cross-section (e.g. at $t=0.75$), $\mathbf{z}^+$ is mostly confined around the discontinuity. To understand this behavior, let us investigate the wave modes which were excited by the driver. An easier way to do this is by looking at the longitudinal cross-section in Figure~\ref{long_cross}.
We distinguish three different characteristic speeds, most easily visible from the first wavefront (at $t=0.3$). There are two wavefronts travelling with different speeds above and below the interface, and a strong `bump' at the interface, which decays away from it, travelling with its own phase speed. The phase speed of this `bump', as well as its decaying nature, leads us to identify it as a surface Alfv\'en wave \citep{1981SoPh...69...27R}. The term `surface Alfv\'en wave' shall be understood here in the broader sense, as we are not exactly in the incompressible limit, much like in the way it can be used to describe kink waves of solar flux tubes \citep{2009A&A...503..213G,2012ApJ...753..111G}. Therefore, the wave driver excites essentially pure Alfv\'en waves above and below the interface, and a surface Alfv\'en wave at the interface. As we have seen in Section~\ref{two}, surface waves present both Els\"{a}sser variables while propagating, explaining why $\mathbf{z}^+$ is confined mostly around the interface. The reason why not only a surface Alfv\'en wave is excited is that the driver is not matching the eigenfunction of a surface Alfv\'en wave. The driver is a spatially independent, uniform forcing, while the surface Alfv\'en wave has an exponentially decaying behaviour away from the interface. The bending of the wavefront seen in Figure~\ref{long_cross} is essentially the linear process of phase mixing \citep{1983A&A...117..220H}, but for the situation in which the inhomogeneity is in the perturbation direction, discussed in, e.g., \citet{1992ApJ...396..297M,1998JGR...10323691G}, and is due to the linear advection term in Eq.~\ref{fullperp}. The linear nature of the process is confirmed by the similar evolution for the two different amplitudes in Figure~\ref{long_cross}. However, the differences seen at later times are revealing: note the fine, small-scale structures evolving in the high amplitude run across the magnetic field, on top of the bent wavefronts. This small-scale generation is nonlinear in nature, and it is confined to the vicinity of the interface. Also note that, while there is a cascade to small scales in the perpendicular direction, the solution remains smooth along the magnetic field, a well-known property of MHD turbulence with a strong guide field \citep{2003matu.book.....B}. 
In Figure~\ref{nonlin-advec}, the linear advection term is plotted against the nonlinear terms. 
These essential nonlinear terms are the well-known nonlinear advective term (last term on LHS of Eq.~\ref{fullperp}) and a nonlinear term that is nonzero only in the presence of density gradients (second term on RHS of Eq.~\ref{Elsasser}, with $\mathbf{v}_A$ a perturbation). For the low amplitude simulation, the nonlinear terms are by far the strongest at the interface, while away from the interface they are negligible. Note the main difference between the two nonlinear terms: the second nonlinear term can only act at the interface, i.e. at the strong density gradient, while the nonlinear advective term is exponentially decaying away from the interface, therefore it is also acting away from it. This leads to the important observation that surface/global Alfv\'en waves are not only inducing nonlinear deformations at the interface, but also act throughout the plasma and can self-cascade away from the interface. In the low amplitude case, the nonlinear term involving the density gradient is clearly dominant over the nonlinear advective term, and is responsible for the deformation of the interface depicted in Figure~\ref{densplot}. Note that the nonlinear term is on average on the order or less than the linear advection term for the low amplitude case. In the high amplitude case, the nonlinearity is initially mostly affecting the interface, but the advective nonlinearity eventually becomes stronger and it appears to affect the whole region around the deformed interface. In this case the nonlinear term is on the order or higher than the linear advection term. Note that in Figure~\ref{nonlin-advec}, the colour bar for the linear advective term is scaled ten-fold between the two runs, while the colour bar for nonlinear terms hundred-fold, in accordance with our expectation of the strength of nonlinearity. The growth of the nonlinear advective term observed in Figure~\ref{nonlin-advec} is expected as $\mathbf{z}^+$ appears to strengthen later in time, as shown in Figure~\ref{Els_hi} and \ref{Els_lo}. It is unclear at this point why the enhancement of $\mathbf{z}^+$ is taking place. However, it appears to be nonlinear in nature, by comparing the low and high amplitude figures. One explanation might be the modification of the eigenfunction of the surface Alfv\'en wave, due to the deformation of the interface. \\
As we mentioned previously, there is a weak numerical reflection of the propagating waves at the upper, open boundary. Therefore, it is crucial to discuss whether the reflected waves are responsible or can modify the turbulent cascade in the simulation. There are several arguments which dismiss the importance of the reflected waves. First of all, the reflection is very weak (maximum $1\%$ reflected wave energy), and therefore the resulting nonlinear advective term must be very small. Indeed, the nonlinear advective term is on the order of $\approx 10^{-6}$ for the low amplitude run, at the $z$-axis boundaries (i.e. furthest from the interface). For comparison, the nonlinear terms in the vicinity of the interface for the low amplitude run are on the order of $\approx 10^{-2}$. Secondly, it is also clear in Figures~\ref{densplot}-\ref{nonlin-advec} that the nonlinearity is mostly affecting the vicinity of the interface, where most of the co-propagating $\mathbf{z}^+$ is situated. If the reflected waves would cause a cascade, this should be distributed in the whole simulation, as the whole boundary is driven equally. Last but not least, the reflected waves have the same polarization as the incident waves, therefore in theory the nonlinear advective term for reflected waves should vanish. Thus, the weak nonlinearity at the $z$-axis boundaries is also originating from the surface wave, which decays exponentially away from the interface. Still, one could argue that the reflected waves could interact with the surface Alfv\'en wave, which has a differently polarized eigenvector. However, the co-propagating $\mathbf{z}^+$ of the surface Alfv\'en wave is on average an order of magnitude larger than the reflected $\mathbf{z}^+$. \\
We have also studied the perpendicular spectra of relevant quantities, such as energy, density, Els\"{a}sser variables, etc. It is clear that higher wavenumbers are being populated in time, however we opted not to show these spectra as the value of the power-law slope might be unreliable. First of all, the available numerical resolution is limiting the extent of the inertial range to about a decade in k-space, restricting the accuracy of the fitting. Secondly, turbulence might not be fully developed yet at the end of the simulation, therefore averaging in time for a smoother spectra is not reliable. Lastly, we have noticed that the perpendicular spectra of the perturbations along the magnetic field is shallower than that of the perpendicular perturbations. As the amplitude of the induced perturbations along the field depend on the plasma beta, it appears that the slope of the energy spectra also depends on plasma beta. This aspect is very interesting and further research is needed to clarify this dependence. 

\section{Discussion and Conclusions}\label{four}  

The underlying mechanism of turbulence generation in MHD is generally accepted to be the nonlinear mutual deformation of colliding, counterpropagating Alfv\'en waves, propagating along some mean magnetic field. This observation is based on the fact that, in the Els\"{a}sser formulation of the MHD equations, the nonlinear advective term requires both Els\"{a}sser variables to be nonzero. In an incompressible and homogeneous plasma, the two Els\"{a}sser variables represent upward and downward-propagating pure Alfv\'en waves of arbitrary amplitude, respectively. Hence, the requirement for both Els\"{a}sser variables to be nonzero simply means the existence of counterpropagating Alfv\'en waves in an incompressible and homogeneous plasma. In this paper we show that once inhomogeneities across the magnetic field are accounted for, new modes appear which have properties different than the waves resulting from a normal mode analysis of a homogeneous plasma. The simplest example of these new modes, described in this work, are the incompressible surface Alfv\'en waves on a planar discontinuity in Alfv\'en speed across the magnetic field. Of course, the discussion can be generalized to arbitrarily inhomogeneous cross-sections, in which case the waves are referred to as global waves. These waves propagate perturbations both along and across the magnetic field, and have linear total pressure perturbations, in contrast to the pure Alfv\'en and pseudo-Alfv\'en waves. Most importantly here, these waves are necessarily described by both Els\"{a}sser variables, while propagating unidirectionally. This should not come as a surprise, as in general once waves are not pure Alfv\'en waves any more, they represent perturbations in both Els\"{a}sser fields. This is valid also for the magnetoacoustic waves in homogeneous, compressible MHD. Therefore, in plasmas which are compressible or inhomogeneous, or both, the requirement for the two Els\"{a}sser variables to be nonzero in order to have nonlinear advection has a different physical meaning: this condition is now satisfied also for unidirectionally propagating waves, without the need for counterpropagating waves. In this sense, waves can self-cascade or self-deform nonlinearly, leading to what we term uniturbulence. The self-cascade of waves can be understood as a coherent nonlinear interaction of Els\"{a}sser variables \citep[for a discussion, see][]{1989PhRvL..63.1807V}. This is an important distinction from the well-known counterpropagating wave scenario, where the interactions are incoherent and act for a shorter period, as the waves interacting are advected in opposite directions. From a theoretical point of view, another important difference from the homogeneous case is the presence of an additional term in the evolution equation of Els\"{a}sser variables, non-zero only when density inhomogeneities are present. This nonlinear term acts together with the nonlinear advective term, causing additional self-deformation. To illustrate the difference between the two nonlinear terms, one can think of a plasma in which inhomogeneities are given by interfaces of sharp change in density, or contact discontinuities. Then, while the nonlinear advection term has a space-filling character, the nonlinear density gradient term mostly deforms the interfaces, in case of global waves present in the plasma. \\
In order to test our prediction that unidirectionally propagating waves can self-cascade nonlineary, we ran full 3D MHD numerical simulations of a simple setup with surface Alfv\'en waves excited, as a proof of concept. By analyzing the results, it is clear that the surface Alfv\'en wave initiates an energy cascade, generating smaller scales, and an overall deformation of the initially planar interface. The self-deformation is initially manifested mostly in the interface via the nonlinear density gradient term, however later on small-scales appear in the vicinity of the interface as well, via the nonlinear advection term. It is unclear whether a steady-state turbulence in the statistical sense is achieved or achievable in such conditions as here. We have also noticed a number of interesting properties of the simulated self-cascade, such as the apparent growth of the relatively smaller Els\"{a}sser variable over time, or the shallower perpendicular spectra of the velocity and magnetic field perturbation along the magnetic field, which could result in a total energy spectra likely depending on the plasma $\beta$. However, as the primary goal of the simulations is a proof of concept, we leave these questions open and as subject for future research. We did not discuss the possibility of equilibrium flows, especially shearing flows along the magnetic field. Even in an otherwise homogeneous plasma, the velocity shear effectively leads to a transverse Alfv\'en speed gradient, and Eqs.~\ref{Elsasser_inhomperp} apply. Therefore, the discussion following Eqs.~\ref{Elsasser_inhomperp} is valid also for an Alfv\'en speed gradient given by a velocity shear. In fact, \citet{2013ApJ...769..142H} studied velocity-shear-induced mode coupling of MHD waves and hinted that the resulting wave behaviour might lead to coherent interactions and to turbulence.\\
The evolution of uniturbulence, presented in Section~\ref{three}, has similarities in appearance to other well-known instabilities relevant for turbulence generation. These are the Rayleigh-Taylor (RT) and Richtmyer-Meshkov (RM) instabilities (I) \citep{ZHOU20171,ZHOU20172}, and the Kelvin-Helmholtz (KH) instability  \citep{1961hhs..book.....C}. First, let us focus on the RT and RM instabilities. These instabilities appear at an interface between two fluids of different densities. The trigger is any misalignment between the (total) pressure gradient and density gradient at the interface. This leads to a non-zero baroclinic term, generating vorticity \citep{2011A&A...526A...5S}. This vorticity can further enhance the misalignment, leading to the instabilities. The pressure gradient exists either because of hydrostatic pressure in the presence of acceleration/gravity (in RTI), or due to a shock wave traversing the interface (in RMI). The misalignment can result from a perturbation to a planar interface with non-zero wavenumber along it, or a non-planar interface. In our simulations without gravity, the driven surface Alfv\'en wave is the only source of acceleration to the interface. Therefore, both the displacement of the interface and the total pressure gradient are the result of the perturbation, meaning that baroclinic terms appear only nonlinearly. This implies that linear surface Alfv\'en waves are not RT-unstable, in agreement with their dispersion relation \citep{1981SoPh...69...27R}. Finite-amplitude surface Alfv\'en waves might still be nonlinearly RT-unstable. Nevertheless, the identification of uniturbulence with nonlinear RTI is not straightforward, as our description in Section~\ref{two} is for a constant density plasma, which by definition is not RT-unstable. Still, the nonlinear density gradient term (second term on RHS of Eq.~\ref{Elsasser}) might be associated with the nonlinear RTI. The possibility of a nonlinear instability strengthens our description that uniturbulence is the result of self-deforming or self-cascading waves. \\
Second, the KHI appears as a result of a velocity shear with an inflection point in a fluid. The surface Alfv\'en wave solution with nonzero perpendicular wavenumber along the interface presents velocity shears at the interface. In our simulation there are no background flows, therefore KHI would also have to appear nonlinearly. However, along with a velocity shear, the surface Alfv\'en wave presents magnetic field perturbations as well, which is known to inhibit the growth of KHI. As such, propagating Alfv\'en waves are known to be stable to KHI \citep{1983A&A...117..220H}. This does not rigorously translate to KH-stable surface Alfv\'en waves, but it hints at stability. The velocity shear in the simulations is less than the root-mean-square of the induced perpendicular Alfv\'en speed along the interface, a condition for KH stability \citep{1961hhs..book.....C}. Of course, this condition is not for time-dependent fields, but a complete KH stability analysis is beyond the scope of this paper. Therefore, with the above arguments we cannot rigorously exclude nonlinear KHI. For related studies of KHI in time-dependent flows see, e.g. \citet{2019ApJ...870..108B,2019MNRAS.482.1143H}.\\
Finally, allowing for a little speculation in this matter, we could say that RTI and KHI arising nonlinearly as it might occur here, could equally well be described as manifestations of uniturbulence. That is, nonlinear, coherent interactions of Els\"{a}sser variables, leading to turbulent flows. This might explain why their morphology seems similar. In summary, the connection between uniturbulence and instabilities is not fully understood, but it is an exciting topic which warrants future research.
\\ 
The analysis presented in Section~\ref{two} is for an incompressible plasma. We chose an incompressible MHD  description as uniturbulence is already present. Hence it can be understood without the complications and extra terms in the equations related to compressibility. However, the solar corona and solar wind are known to be slightly compressible \citep{2013LRSP...10....2B}, and in some cases the nearly-incompressible \citep{1992JGR....9717189Z,1993PhFlA...5..257Z}, or fully compressible description may need to be considered. The description of uniturbulence in a compressible plasma is left for future work.
Uniturbulence might likely be relevant in plasmas structured across the magnetic field, such as the solar corona and solar wind. Specifically, it could provide an additional cascade rate and therefore enhanced dissipation, especially in open magnetic field regions. Nevertheless, having in mind the coherent nature of the interaction, it might be relevant even in closed magnetic environments, such as coronal loops. Again, future research is needed to determine the importance of uniturbulence as compared to the classical counterpropagating cascade in different plasma environments.  

\begin{acknowledgements} T.V.D. was supported by the GOA-2015-014 (KU Leuven) and the European Research Council (ERC) under the European Union's Horizon 2020 research and innovation programme (grant agreement No. 724326). \end{acknowledgements}

\bibliographystyle{apj} 
\bibliography{../Biblio}{} 

\end{document}